\documentclass{PoS}

\title{AGN jet models}

\ShortTitle{AGN jet models}

\author{\speaker{Christian R. Kaiser}\\
        School of Physics \& Astronomy, University of Southampton, Southampton, SO17 1BJ, UK\\
        E-mail: \email{crk@soton.ac.uk}}

\abstract{In this review I concentrate exclusively on models for the large-scale structure created by jet flows in AGN. I briefly mention models for the evolution and emission of these objects and how they can also be applied to microquasars. While in radio-loud AGN we can directly use the radio synchrotron emission of these structures, we need to find other detection methods in microquasars. Where possible, the application of AGN models has produced important insights into the time-averaged energy transport rate of microquasar jets. I also describe methods for using the large-scale structure of jets to infer jet duty cycles. Finally, I point out some recent work taking the idea of a connection of accretion disc states and jet production from microquasars and applying it to radio-loud AGN.}

\FullConference{VI Microquasar Workshop: Microquasars and Beyond\\
		September 18-22 2006\\
		Societ\`a del Casino, Como, Italy}

\begin{document}

\section{Introduction}

Since the first discovery of Galactic microquasars, hopes were very high that we might learn something from them to explain the AGN phenomenon. Of course the opposite direction was also suggested and a number of ideas first developed for AGN have been applied successful to microquasars, not least their names. This field of study, the crossing between the Galactic and extragalactic boundary, has exploded over recent years. The classical book `Beams and jets in astrophysics', edited by Philip Hughes and published in 1991, contained only 80 out of a total of 560 pages on Galactic jet sources. In fact, most of these 80 pages discussed the jets of young stellar objects. SS433 gets just over eight pages, while jets from X ray binaries are summarised within less than one page. I doubt that, were this book re-written today, this particular balance would still be applied. 

There is a lot the Galactic and extragalactic jet communities can learn from each other. Below I am trying to collect together some of the more theoretical issues surrounding the large-scale structure created by AGN jets and how these may apply to the microquasar jets. I find it impossible though to stick to this one way view and so I will also briefly discuss some work that crosses the gulf the other way. 

In section \ref{fr} I present the great division in the morphology of the large-scale structure of AGN jets. I go on to discuss theoretical models for both classes, with a very heavy bias towards the structure of the more powerful jets, in section \ref{mods}. I show the application of these models to microquasars in section \ref{appl}. Jet duty cycles are discussed in section \ref{duty}. In section \ref{revis} I outline some recent work on AGN jets that takes a leaf out of the book of microquasars in terms of the association of accretion states with jet production. A very brief summary follows in section \ref{conc}.

\section{The two FR classes}
\label{fr}

The large-scale radio structure of radio galaxies and radio-loud quasars falls into one of two distinct classes \cite{fr74}. The Fanaroff-Riley type I (FRI) objects are the less luminous and their surface brightness decreases from the centre outwards. The more luminous FRII sources show an edge-brightened morphology. The division between the two classes lies at a luminosity of about $5\times 10^{25}$\,W\,Hz$^{-1}$ at 178\,MHz. However, there is evidence that this luminosity depends on the optical luminosity of the host galaxy of the AGN as $L_{\rm B}^{1.8}$ \cite{lo96}.

High resolution radio observations of FRI sources show collimated, presumably laminar jet flows on scales of 10s, in many cases 100s, of kpc. The jets end in very luminous radio `hotspots' and are surrounded by diffuse radio lobes of much lower surface brightness. The presence of two bright radio hotspots in opposite directions from the central host galaxy has led to the name `classical double' radio source. 

Sources in the FRI class can be further distinguished into sources with well-defined, roughly ellipsoidal lobes similar to those of FRII-type sources, but without prominent hotspots (`fat doubles'). The other sub-class of FRI objects contains jet flows with a bright `flare point' close to the centre of the structure and broad, turbulent jets further out which fade gradually as they propagate outwards. The structure of these jets is reminiscent of the turbulent flow out of a chimney. 

Apart from the radio morphology there also also other differences between the properties of the two FR classes. Usually the optical line emission is weaker in FRI-type objects \cite{hl79,zb95}. The optical continuum emission of the AGN in FRI objects correlates with the luminosity of their radio cores while the optical luminosity of FRII sources exceeds the value predicted by this correlation for a given radio luminosity \cite{ccc00}. This points to a radiatively inefficient accretion regime in the AGN of FRI-type sources compared to a more efficient state of FRII objects \cite{fkm04}. Further support for such an interpretation comes from X ray observations \cite{hec06} and the infrared emission of the dust in the host galaxies \cite{mhs04}.

In the following we will first concentrate on models for the large-scale radio structure and how they may be applied to microquasars. We will then return to the other differences between the FR classes and briefly discuss how the results obtained for microquasars may be used to unify the radio-loud AGN.

\section{Models for the large-scale radio structure}
\label{mods}

\subsection{Models for FRI sources}

The turbulent jet flows of FRI-type sources are notoriously difficult to model. Numerical models can be used to study the onset and evolution of fluid instabilities leading to the turbulent disruption of the jet \cite{ph04}. However, the fastest growing instability modes are those on the smallest physical scales. Therefore studies of turbulent jets are confined to small sections of the jet flow and global simulations of the jet dynamics, let alone the emission properties of the jet material, are beyond the current capabilities.  

Analytical models for these turbulent jets are equally difficult to construct. Some insight can be gained from the application of conservation laws to the jet flow \cite{gb94,gb95}. Also, models for the velocity field, magnetic field structure and distribution of relativistic electrons can be crafted onto high resolution radio observations of the turbulent jets of FRI-type objects \cite{lb02a,lb02b,lb04}. Unfortunately it is not clear how such models can be generalised to make predictions for the dynamics and emission properties of the FRI class as a whole.

In the absence of a good, general model for the dynamics and emission properties of turbulent jet flows, we can only summarise a few basic facts. Turbulent jets in general give rise to much fainter large-scale radio structures than the laminar jet flows we will discuss next. The main reason is the absence of an efficient acceleration site for relativistic electrons like a strong shock which most of the jet material has to pass through. Therefore, if the conditions in microquasars are such that they usually develop turbulent jets, then we may not be able to easily detect their large-scale structure.

\subsection{Models for FRII sources}

The situation is much better for FRII-type structures. Observations indicate laminar flow in the jets of these sources and so spatial resolution in numerical simulations is a far smaller problem. It is now possible to simulate the entire jet-lobe structure with fully three dimensional, relativistic MHD codes that also allow us to follow the advection of relativistic electrons responsible for the emission of synchrotron radiation \cite{tjr02,amg03}. 

While numerical models are invaluable in testing our ideas about the physics in these sources, they can always only explore a small corner of parameter space. Fortunately, the structure of FRII-type jets and lobes can also be described with analytical models.

\subsubsection{Analytical models for FRII sources -- Dynamics}

There is a large number of models describing the dynamics of the jets and lobes of FRII-type objects \cite{ps74,br74,bc89,sf91,ka96b,kc97,pa00}. Almost all models are based on the idea of laminar jets ending in strong shocks identical to the observed radio hotspots. After passing through the shock, the jet material inflates the lobe which is overpressured with respect to the surrounding medium and so drives a bow shock into this material. The jet is forced into pressure equilibrium with its own lobe. Thus the lobe ensures the collimation of the jet flow. It also protects the jet from contact with the high density environment which would cause the entrainment of heavy material into the jet flow and so the disruption of the jet by turbulence. The shock at the end of the laminar jet is most likely the site of efficient acceleration of the relativistic electrons which then give rise to the radio synchrotron emission of the lobe \cite{hr74}.

It is interesting to note that the reconfinement or collimation of the jet inside the lobe will only occur if the pressure inside the jet falls below the pressure in the jet environment. For a ballistic jet the pressure of the jet material will decrease as $r^{-2}$, where $r$ is measured along the jet axis. If the pressure in the jet environment decreases as $r^{-\beta}$ with $\beta > 2$, then the jet never needs to adapt to its surroundings and will remain ballistic \cite{sf91}. In this case, it is very likely that the jet flow never develops a strong shock at which electrons could be accelerated. Should this situation occur in microquasar jets, then it will be virtually impossible to observe them on large scales. However, the steep density gradients required for the source environments are unlikely to occur frequently in the ISM.

\subsubsection{Analytical models for FRII sources -- Emission}

Once we have a model for the source dynamics, we can hope to develop a description for the luminosity evolution as a function of the size of the jet-lobe structure or jet age. Most of the relativistic electrons will be accelerated in the region of the radio hotspots and are then advected by the plasma flow into the lobe. There is little evidence for re-acceleration of the electrons inside the lobes \cite{cpdl91}. Energy losses of the electrons due to the adiabatic expansion of the lobe, the emission of synchrotron radiation and the inverse Compton scattering of photons of the cosmic microwave background radiation lead to the development of a break in the electron energy distribution at high energies. This break translates into a steepening of the radio spectrum at high frequencies. The radiative lifetime of the electrons is of order $10^{6 \rightarrow 7}$\,years. Therefore the position of the spectral break can be used to determine a `spectral age' of the lobe \cite{pa87,al87}. The spectral age may not represent the true age of the source structure if the plasma injected into the lobes at different times is well mixed or if the magnetic field in the lobe is inhomogeneous \cite{pt93,pt94,emw97,ck05}.

Returning to the development of an analytical model for the FRII-type sources, we note the need to include these energy losses in our calculations since the timescales associated with them are certainly comparable to, if not shorter than, the typical ages of AGN jets. In the case of microquasars the need to include radiative energy losses may well be reduced because the strength of the magnetic fields expected for their lobes imply radiative lifetimes for the electrons not too different from those in AGN. Several models implement self-consistent treatments of the energy losses into the basic dynamical models for FRII sources \cite{kda97a,brw99,mk02}. All models predict a moderate, negative luminosity evolution for young sources which significantly accelerates as inverse Compton losses become important.

These kind of models allow us to determine the properties of the jets and their environments from directly observable parameters. Typically we need to determine the luminosity and length of the jet from observations. Application of the models then allows the derivation of source ages, the jet power and the density of the gas in the source environment. The source age can be supplemented with a spectral age if observations at more than a single radio frequency are available. In this way, the observed lobes provide us with information on the energy transport rate of the jets. They act as calorimeters of the jet activity.

\section{Application to microquasars}
\label{appl}

Obviously we would like to apply the analytical models outlined above to the radio synchrotron emitting lobes around microquasar jets. Unfortunately, there are very few of these detected. The reason could be that the jets of microquasars are disrupted by turbulence and only develop not very luminous large-scale structures akin to the FRI-type objects. This possibility remains unexplored at present. 

If the jets of microquasars give rise to large-scale structures similar to those observed for FRII-type AGN, then the models outlined above should allow us to determine the expected surface brightness of these lobes. It emerges from such an application of the AGN models that the combination of the microquasar jet power and the density of the ISM is not favourable for a high surface brightness of the lobe in the radio \cite[and Heinz in these proceedings]{sh03}.

We can still apply the models for FRII-type structures, but we have to use observations of other source components. The bow shock surrounding the lobe compresses and heats the gas in the immediate source environment. For AGN, this is the hot ($\sim 10^7$\,K) plasma of the intracluster medium. In nearby FRII sources we can detect the enhancement of the bremsstrahlung emission caused by the action of the bow shock in X rays \cite{swa02}. In the case of the ISM surrounding the lobes of microquasars the gas temperature is much lower. If the bow shock is strong enough, then it may partially ionise the gas and cause the emission of radio bremsstrahlung \cite{kgb04}. We also expect to observe some emission lines from the shocked gas. Both signatures have recently been detected around the postulated lobe inflated by one of the jets of Cyg X-1 \cite[and Russell in these proceedings]{gfk05}.

Using the observational signatures of bow shocks allows us to estimate the time-averaged jet powers of microquasars in the same way as for radio-loud AGN. The results show that the kinetic energy transport rate of microquasar jets is comparable to the bolometric luminosity of the accretion disc \cite[and Gallo in these proceedings]{gfk05}.

\section{Jet duty cycles}
\label{duty}

The evolutionary timescales of microquasar jets are humanly accessible. Hence it is obvious from observations that the jet activity in these systems is not permanent. During the radiatively inefficient accretion state, a moderately powerful, but comparatively stable jet flow is produced, while very powerful, but short lived jet ejection are associated with the transition to the radiatively efficient accretion state \cite[and Fender in these proceedings]{fbg04}.

For radio-loud AGN the activity cycles of the jets are too long to be observed directly. However, the large-scale radio structures can again help us. Their relatively long evolutionary timescales effectively turn them into a long term memory of previous jet activity. The observation of `bubbles' of radio emission displaced from the currently active radio-loud AGN \cite{oek00} have been interpreted as the buoyantly rising remnants of a previous activity cycle \cite{cbkbf00}. In the meantime, many such bubbles have been observed, a large fraction of them only as depressions in the X ray surface brightness in the hot AGN environment [Churazov in these proceedings]. 

The buoyant bubbles are most often associated with relatively weak AGN like M87 or Perseus A. However, in some more powerful radio sources we also find evidence for repeated jet activity. Some FRII-type objects show four, rather than the more common two, lobes. The X-shaped or winged FRII objects are one example \cite{dbl02}. In these particular objects the observational evidence is still inconclusive, because the radiative age of the four lobes does not indicate clearly that two lobes are significantly older than the other two. 

A more convincing example for restarted jet activity is provided by the `double-double radio galaxies' (DDRG) \cite{sbrlk00}. In these sources a pair of outer lobes without signs of currently active jets is aligned with a pair of inner lobes which appear to be supplied with energy by jets. The radio observations are consistent with restarted jet activity on timescales of $10^{7\rightarrow 8}$\,years \cite{ksr00}. In this context it is also interesting to note the recent discovery of a triple-double radio galaxy which contains three pairs of aligned radio lobes [Brocksopp et al., in preparation].

\section{The FR divide revisited}
\label{revis}

The main aim of this review has been to highlight the techniques for modelling the large-scale structure caused by jets from AGN and how such models can be applied to microquasars. I will now briefly point out some recent work that uses results for microquasars as an inspiration for explaining the FR divide in radio-loud AGN.

Using standard models for the dynamics of jet flows, we were able to show that the turbulent disruption of jets depends strongly on the jet power. A secondary role is played by the density of the source environment [Kaiser \& Best, in preparation]. Powerful jets are less susceptible to the development of turbulence than weaker jets. If we now assume that, analogous to microquasars, in AGN weaker jets are associated with radiatively inefficient accretion flows, then we would expect that `optically dull' AGN preferentially develop an FRI-type structure. This is indeed observed. However, because of the influence of the source surroundings, we would expect to see some objects that connect radiatively inefficient AGNs with FRII-type radio structure in exceptionally rarefied environments. These objects may be identified with the LERGs mentioned above. Examples of the opposite combination of a radiatively efficient AGN giving rise to an FRI-type radio structures are apparently rare. This suggests that the jet power associated with the FR divide is higher in typical source environments than the jet power at the transition from radiatively efficient to inefficient accretion. 

If this analogy holds, then the more powerful jets inflating lobes of type FRII are probably associated with the transition to radiatively efficient accretion scheme. In microquasars the jet ejections appearing during this transition last for a few hours while the jet flows in AGN are stable for $10^8$\,years. The ratio implies a scaling of the jet ejection timescale with $M_{\rm BH}^{3/2}$, where $M_{\rm BH}$. Interestingly this scaling is the same as that of the viscous timescale in the accretion disc \cite{fkr92}. 

Clearly the scenario outlined above is quite speculative and needs firm observational support. However there are clear indications of a connection between the jet activity in AGN and their accretion state 
\cite[and Jester in these proceedings]{kjf06}.

\section{Summary}
\label{conc}

In this review I have tried to collect together information on modelling the large-scale radio structure of radio-loud AGN. The main purpose was to show how these well developed models can be applied to microquasar jets. 

We can apply models for the evolution of lobes with FRII-type morphology to the jets in microquasars. The main problem here is the small number of lobes in microquasars that are detected directly because of their radio synchrotron emission. Alternative ways of finding the large-scale structures inflated by the jets include the search for radio bremsstrahlung and/or optical emission lines from the compressed ISM. Where such observations are possible, the AGN models provide valuable constraints on the kinetic energy transported by the microquasar jets. The possibility remains that microquasar jets become turbulent as well in which case the detection of their large-scale structure is complicated further. 

We can also work backwards in the sense of applying the suggested connection between accretion disc states and jet ejections in microquasars to AGN. Some evidence suggests that the low radiative efficiency of AGN with FRI-type morphology may be caused by the relatively weak jets associated with radiatively inefficient accretion states. More evidence is needed to firmly establish accretion states for AGN systems before this suggestion can be investigated in detail.

\bibliography{crk}
\bibliographystyle{PoS}

\end{document}